\documentclass[aps,superscriptaddress,showpacs,nofootinbib,twocolumn]{revtex4}

\input{epsf}

\begin{document}

\title{
A new improved optimization of perturbation theory: applications to \\
 the oscillator
energy levels and Bose-Einstein  critical temperature\\}

\author{Jean-Lo\"{\i}c Kneur}
\affiliation{Laboratoire de Physique Math\'{e}matique et Th\'{e}orique - CNRS -
 UMR
5825 Universit\'{e} Montpellier II, France}

\author{Andr\'e Neveu}
\affiliation{Laboratoire de Physique Math\'{e}matique et Th\'{e}orique - CNRS -
 UMR
5825 Universit\'{e} Montpellier II, France}

\author{Marcus B. Pinto}
\affiliation{Departamento de F\'{\i}sica,
Universidade Federal de Santa Catarina,
88040-900 Florian\'{o}polis, SC, Brazil}

\begin{abstract}
Improving perturbation theory via a variational optimization has
generally produced in higher orders an embarrassingly large
set of solutions, most of them unphysical (complex). We introduce
an extension of the optimized perturbation method which leads to a 
drastic reduction of the number of acceptable solutions.
The properties of this new method 
are studied and it is then  applied to 
the calculation of relevant quantities in 
different $\phi^4$ models, such as the 
anharmonic oscillator energy levels and the critical
Bose-Einstein Condensation temperature shift $\Delta T_c$
recently investigated by various
authors. Our present estimates of $\Delta T_c$, 
incorporating the most recently available six and seven  loop 
perturbative information, are in excellent agreement with
all the available lattice numerical simulations. This
represents a very substantial improvement over previous treatments.

\end{abstract}

\pacs{03.75.Fi, 05.30.Jp, 11.10.Wx, 12.38.Cy}

\maketitle

The variationally improved or optimized
perturbation based on the
linear $\delta$ expansion (LDE)\cite{delta,pms,odm}
is a well-used modification of the usual perturbation theory, based
on a reorganization of the interacting Lagrangian such that it depends on
an arbitrary mass parameter, to be fixed by some optimization
prescription. 
In $D=1$ theories, such as the quantum mechanical anharmonic
oscillator \cite{ao}, the LDE turns out to be  equivalent \cite{deltaconv}
to the ``order-dependent mapping" (ODM) resummation method \cite{odm}. At the
same time, the principle of minimal sensitivity (PMS) \cite{pms}
optimization, which takes  extrema with respect to the
mass parameter, is  equivalent at large orders 
to a rescaling of the adjustable oscillator
mass with perturbative order, which can essentially suppress the
factorial large order behavior of ordinary perturbative coefficients. 
This appropriate rescaling of the adjustable mass 
gives a convergent series \cite{deltaconv,deltac}
e.g. for the oscillator energy levels \cite{ao} and related quantities.
Any physical quantity whose ordinary perturbative sequence is available can then
be evaluated to an order $\delta^k$ using simply modified Feynman rules 
as implied by the following formal substitution valid for a
scalar field theory: 
\begin{equation}
\omega \to \omega\,(1-\delta)^{1/2}\;;\;\;\; g\to g\, \delta \;\;,
\label{subst}
\end{equation}
where $\omega$ and $g$ are the mass and coupling respectively. 
Note that for the $D=1$ quantum mechanical anharmonic
oscillator described by a $g \phi^4$ theory no renormalization
is needed \cite{ao}, while for $D>1$ models, the parameters $g$ and $\omega$
in Eq. (\ref{subst}) are to be considered implicitly bare parameters, and the
procedure can be made in this case fully consistent \cite{gn2,marcus1} with the
renormalization program of ordinary perturbation theory. In particular,
appropriate renormalization takes into account properly any (mass or field)
anomalous dimensions when the latter are relevant.

Now, a definite drawback of the optimization prescription is that it involves
minimization of a polynomial equation of order $k$ in the relevant
mass parameter $\omega$ at perturbative order $\delta^k$, such that more and
more solutions, most of them being complex, are to be considered when 
increasing the order. This non-uniqueness of the optimized solution 
requires extra choice criteria, and thus may
seriously obscure  the interpretation and 
the convergence towards the correct result in many
non-trivial cases where the exact non-perturbative result is totally
unknown. Moreover, the fact that most  solutions are complex
is  embarrassing, as one has to invoke still an extra 
criterion  to select a (supposedly correct) real result.
 For some of the simplest models where the method applies, like
 the oscillator energy levels typically, fortunately
all of the complex optimization solutions have actually 
small imaginary parts and rapidly decreasing
as the perturbative order increases (see e.g. Bellet et al in refs. 
\cite{deltac}), so that the convergence properties are not much
obscured by this inconvenience of the PMS. But in less trivial
situations, the imaginary parts of the PMS solutions may be 
large (see e.g \cite{pra,new,braaten}) and 
thus their physical intepretation unclear.

In this paper, we propose a simple generalization
of the PMS criterion as performed on the LDE series, which turns out to
lead to a drastic reduction of physically acceptable
real optimization solutions at each successive perturbative order
in all the physical cases we have applied it to. First, we treat
 the oscillator ground state energy, both in the large-$N$
case (for the vector $O(N)$-symmetric $\phi^4$ 
model) and the ordinary oscillator (scalar, $N=1$) $\phi^4$ model.
We then  apply it to a less trivial and  
more interesting problem associated with the breakdown of perturbation
theory near a critical point, namely  the evaluation of the critical
transition temperature for a dilute, weakly interacting homogeneous Bose gas.
This has been the source of
controversy for many years and recently several independent groups 
have provided comparable evaluations of the critical temperature.
 The relevant
field theoretic framework is  a
$\phi^4$ $D=3$ model
(after dimensional reduction) with an $O(2)$ symmetry (see e.g.
Refs.\cite{zustin,second,baymprl,baymN,arnold1} for reviews). Here, for
completeness we also consider the  large-$N$ limit of the $O(N)$ symmetric
model case where the exact  next-to-leading $1/N$ result is known analytically.
In all these cases, the new method seems to give excellent approximations
 in comparison with the standard PMS ones.

\section{basic method and the oscillator energy levels}
Let us start with the basic perturbative series of the 
oscillator ground state energy level as described by a
$g\phi^4$  $D=1$ model with mass $\omega$:
\begin{equation}
E^{(n)}_0 =  \frac{\omega}{2} +\omega 
\sum^n_{q=1} (-1)^{q+1}\:c_q\:
\left (\frac{g}{\omega^3}\right )^q \;\;,
\label{Eaopert}
\end{equation}
with  $c_1=3/4$, $c_2 = 21/8$, etc\ldots\cite{ao}. 
As discussed in the introduction, the LDE procedure is implemented 
by the substitutions
\begin{equation}
\omega \to \omega (1-\delta)^{1/2}\;,\;\;\; g\to g\,\delta\;,
\label{subst1}
\end{equation}
into the perturbative series Eq. (\ref{Eaopert}) then reexpanding the latter
to order $k$ in the new expansion parameter $\delta$. Next, $\delta$ is set
to the value $\delta=1$, such as to recover the original (massless) theory,
while at any finite order $k$ there remains a
dependence on $\omega$ in the LDE result $E^{(k)}(\omega,\delta=1)$. 
The standard, mostly used optimization criterion is the
principle of minimal sensitivity
(PMS), requiring at each successive perturbative orders $k$: $\partial
E^{(k)}(\omega,\delta=1)/\partial\omega \equiv 0$ for optimal $\omega$ values.
The modification (generalization) that we propose here is first to introduce
extra arbitrary parameters, starting at order two with one more parameter:
\begin{equation}
\omega \to \omega\, (1 - \delta)^{1/2} [ 1 + (1-a)\, \delta ]^{1/2}\;\;,
\label{subst2}
\end{equation}
such that the modified Lagrangian
still interpolates between the free field (massive) theory for
$\delta=0$ and the original (massless) theory for $\delta=1$.
Second, now the PMS criterion is generalized by requiring  
both $\partial E^{(2)}/\partial \omega =0$,
and $\partial^2 E^{(2)}/\partial \omega^2 =0$, which gives a system
of two equations to be solved simultaneously for $a$ and $\omega$:
For $a=1$,  $E^{(2)}(\omega)$ has no real minimum, but an inflection
point with an almost horizontal tangent, and a small value of $a-1$
makes this tangent horizontal, removing the embarrassment of complex
extrema.
At higher orders, the generalization is easily done with additional
parameters and additional vanishing of higher derivatives of 
$E^{(k)}(\omega)$.
For example, at third order, we introduce two parameters 
($a$ and  $b$) as
\begin{equation}
\omega \to \omega\, (1 - \delta)^{1/2} [ 1 + (1-a)\, \delta + b\,
\delta^2]^{1/2} \;\;, 
\label{subst3}
\end{equation}
together with the extra requirement on the third derivative:
 $\partial^3 E^{(3)}/\partial \omega^3 =0$\footnote{For 
completeness, note that we generalize
Eq. (\ref{subst3}) at arbitrary higher orders by
simply adding to the expression 
within the right-handed bracket of Eq. (\ref{subst3}), 
a term $b_k \delta^k$ at each order $k$, where $b_k$
are new parameters.}.

The results of the new method for the (scalar case) oscillator energy level
are shown in Table I, up to order nine, where
higher order generalizations of Eq. (\ref{subst3}) are
still reasonably tractable upon
using efficient polynomial equation solver (we used mainly
{\it Mathematica} \cite{mathematica})\footnote{It is clearly
becoming more cumbersome and CPU consuming
to apply such a method at very large perturbative orders due to the
increase in the number of parameters at successive orders.}.  
The real solution is unique (at least 
up to the highest perturbative orders that we
analyzed, as given in Table I) 
and we also indicate for
later discussion the corresponding values obtained 
for the extra interpolation 
parameter $a$ as defined in Eqs. (\ref{subst2}), (\ref{subst3}). (Note
that actually the (unique) real optimization solution gives values for the 
other parameters $b$, \ldots
in e.g. Eq. (\ref{subst3}) at higher orders, which are
almost always small with respect to $a$, more
on this later).

We see that after approaching the exact answer to within $4 \times 10^{-3}$ at
fourth order, the approximation becomes worse in a way strongly reminiscent
of the behavior of an asymptotic expansion with alternating signs
evaluated at a finite value 
of its parameter. This is clearly related to the
fact that the original series in Eq.~(\ref{Eaopert}) 
has factorially growing coefficients at large orders,
and thus our alternative method 
apparently  looses the property of compensating this divergent behavior by an
appropriate rescaling\cite{deltaconv,deltac}
of the optimized mass parameter, unlike the standard
LDE-PMS. More precisely, one can see from the third column of Table I
that the values of the optimized mass parameter $\tilde \omega$,
 after increasing regularly with the perturbative order for the
first few orders, appear rather to have a slower increase for the highest 
perturbative orders we could consider. 

Nevertheless, we can try to
better exploit these results by standard resummation methods applied on
the obtained perturbative sequence, typically
considering simple Pad\'e approximants.
Accordingly, let us define the following perturbative series:
\begin{equation}
E^{(p)} = E_1 + \sum_{q=1}^p (E_{q+1}-E_q) x^q \;\;,
\label{padseries}
\end{equation}
where $E_q$ denotes the results for the ground-state energy at orders $q$ in
Table I, so that Eq.~(\ref{padseries}) simply gives $E^{(p)} = E_p$ for $x
=1$. Now a simple $(3,3)$ Pad\'e approximant of 
Eq.~(\ref{padseries}) using the first seven orders of  Table I 
gives $0.420841$,
and a $(4,4)$ Pad\'e using all the numbers of Table I gives $0.420838$, still
an improvement,  within $8\times 10^{-5}$ of the exact answer, in spite of the
superficially disastrous results of  the eighth and nineth orders.\\
\begin{table}
\begin{center}
\caption{Improved-PMS results for the oscillator ground state energy level, 
at different orders $k$, with the corresponding values of 
the optimal mass parameter $\tilde \omega$ and the main interpolation
parameter $a$. 
$E_{exact}= 0.42080497 \ldots (4\,g)^{1/3} $. }
\begin{tabular}{|c||c|c|c|}
\hline
$ k $ &  $E_{IPMS}/(4\,g)^{1/3}$ & $\tilde \omega$ & $a$ \\
\hline\hline
1   &  0.429268   & 1.82  &   \\
\hline
2   &  0.418483 & 2.04    & 1.05  \\
\hline
3  &   0.422341 &  2.18  & 1.08    \\
\hline
4 & 0.419138 &  2.28 &  1.09  \\
\hline
5 &  0.423496 &  2.37 &  1.11 \\
\hline
6 & 0.41523 & 2.45 &  1.12 \\
\hline
7 & 0.436015 & 2.51 &  1.15 \\
\hline
8 & 0.380812  & 2.59 &  1.15\\
\hline
9 & 0.647259  & 2.60  &  1.25 \\
\hline
\end{tabular}
\end{center}
\end{table}
Next, the same analysis is performed for the large-$N$ approximation of the
vector $O(N)$ symmetric oscillator, which consists of considering only the
``cactus" Feynman graphs. As is well-known, 
the resulting perturbative series for the ground state energy 
of a form similar
to Eq.~(\ref{Eaopert})
can be obtained simply in this case to arbitrary orders by 
solving the large-$N$ gap equation exactly,
and expanding in a 
perturbative series in $g$ the corresponding exact 
large-$N$ expression of the
ground state energy.  
When written with the same normalization
as in Eq. (\ref{Eaopert}) the first few coefficients of this series
are $c_1=1/4$, $c_2 = 1/4$, $c_3=1/2$,\ldots.
However, an important difference with the scalar case oscillator
 is that the
resulting  perturbative series has no factorially growing coefficients 
at large orders and a finite convergence radius. 
As one can see, the method performs very well in this case, approaching
rapidly very close to the exact solution when the order increases.
Again, there is only one real positive solution at each order.
One can also notice from the
last column of Table II the more regular behavior of the extra 
interpolation parameter $a$, as compared to Table I, which now appears
to tend towards a constant value. (The other 
parameters $b$, etc\ldots corresponding to the analysis in Table II
are numerically quite negligible with respect to $a$, so that we do not give 
their explicit values).

In order to better understand 
why the method seems not so efficient when applied
to a factorially divergent series, like Eq. (\ref{Eaopert})
for the standard (scalar case) oscillator, 
let us consider now
the even simpler functional integral of euclidian $\phi^4$
theory in zero dimension:
\begin{equation}
I(g,m) \equiv \frac{\sqrt{2}}{\Gamma(1/2)}\:\int_0^\infty dx \exp\{-m^2\,x^2/2
-g\,x^4/4\} \,,
\end{equation}
As is well-known, this can be expanded in a perturbative series in $g$
with alternating signs, factorially growing coefficients at large orders:
\begin{equation}
I^{pert}(g,m) = \frac{1}{\Gamma(1/2)\,m}\;\sum_{p=0}^\infty
(-1)^p \frac{\Gamma(2p+1/2)}{p!}\:\left (\frac{g}{m^4} \right )^p
\end{equation}
Now the exact result of the integral for $m=0$ is
\begin{equation}
I(g,0) = 2^{-
\frac{3}{2}} \frac{\sqrt{2}\Gamma(1/4)}{\Gamma(1/2)}\: g^{-1/4} =
 1.0227656721\ldots \: g^{-1/4}\;\;,
\end{equation}
which we can compare with the results of the application of the method: 
$0.954325$, $1.0865771$, $0.94213824$, $1.244391803$, $0.7038846$
respectively for orders one to five. Then, there is no real solution at
order six. Nevertheless, a Pad\'e approximant 
$P_{[1,1]}$ (thus constructed from the first three
orders) still gives $1.01754$. 
Thus, the behavior is quite similar to the one of 
the (scalar) oscillator where there are also no physically acceptable 
solutions at higher orders. (However, the quality of the results
from Pad\'e approximants of higher orders appears  
 to deteriorate more rapidly than in the previous oscillator case:
 For example, a  $P_{[2,2]}$ thus constructed from the first five
orders gives $1.05677$). 
This anyway confirms that the generalized method is not
appropriate to turn a factorially divergent large order behavior
into a more convergent one, while it definitely improves the 
LDE convergence when starting from a less divergent original series. 
\begin{table}
\begin{center}
\caption{Improved-PMS results for the  oscillator ground state
energy level at large $N$,  for different orders $k$, with the corresponding
values of  the optimal mass parameter $\tilde \omega$ and
interpolation parameter $a$. $E_{exact}=
0.4292678409\ldots (4\,g/3)^{1/3}$.} 
\begin{tabular}{|c||c|c|c|}
\hline
$ k $ &$E_{IPMS}/(4\,g/3)^{1/3} $ &  $ \tilde \omega $ & a \\
\hline\hline
1   &  0.429268 & 1.26    &    \\
\hline
2   &  0.429589 & 1.37   & 0.96   \\
\hline
3  &  0.429400 &  1.43 & 0.92      \\
\hline
4 &  0.429326 &  1.47  &  0.89  \\
\hline
5 &  0.429296  &  1.50 & 0.87 \\
\hline
6 & 0.429282 &  1.52  &  0.85  \\
\hline
7 & 0.429275  & 1.54  &  0.84 \\
\hline
8 & 0.429272  & 1.55   & 0.83 \\
\hline
9 & 0.429270   &  1.56  &  0.82 \\
\hline
\end{tabular}
\end{center}
\end{table}

\section{Critical theory and BEC critical temperature}
We now turn to the more recent and challenging problem
of the BEC critical temperature evaluation. 
We first recall that, to $O(a^2 n^{2/3})$,  its  functional form was found to be
\cite {second}
\begin{equation}
T_c=
T_0 \{ 1 + c_1 a n^{1/3} + [ c_2^{\prime} \ln(a n^{1/3}) +c_2^{\prime \prime}
] a^2 n^{2/3} \} \;\;,
\label{Tcexp}
\end{equation}
where $T_0$ is the ideal gas condensation
temperature, $a$ is the $s$-wave scattering length, $n$ is the density and
$c_1, c_2^{\prime}, c_2^{\prime\prime}$ are numerical coefficients. Some early
analytical predictions included the self-consistent
resummation schemes \cite {baymprl} ($c_1\simeq 2.90$), the $1/N$ expansion at
leading order \cite{baymN} ($c_1\simeq 2.33$) and at next to leading order
\cite{arnold1} ($c_1\simeq 1.71$) and also the  LDE
 at second order \cite{prb} ($c_1\simeq 3.06$). The numerical methods include
essentially lattice simulations (LS). The most recent LS
results are reported by the authors of Ref. \cite{russos} ($c_1 = 1.29
\pm 0.05$) and of Ref. \cite{arnold2} ($c_1 = 1.32 \pm 0.02$). Very recent
analytical studies predict $c_1= 1.27 \pm 0.11$ \cite {kastening2}. 
The problem is that these 
coefficients (except
$c_2^\prime$) are sensitive to the infrared
physics at the critical point and so, no
perturbative approach can be used to compute them.
At the critical point, one can describe a weakly interacting dilute
homogeneous Bose gas by an effective action analogous to a 
$O(2)$ scalar field model in three-dimensions given by

\begin{equation}
S_{\phi}=  \int d^3x \left ( \frac {1}{2} ( \nabla \phi )^2 +
\frac {1}{2} \omega
\phi^2 + \frac {g}{4!} \phi^4
\right ) \;,
\label{action2}
\end{equation}
where $\phi$ is a two-component real scalar field.
The parameters $\omega$ and $g$ are related to the original
parameters of the nonrelativistic action by
$\omega=-2m\mu$ and $g=48 \pi a mT$ \cite{baymN,arnold1} 
where $\mu$ represents the chemical
potential, $m$ the atomic mass and $T$ the temperature.
The leading order coefficient of the critical temperature
shift can be expressed as \cite{baymprl} 
\begin{equation}
c_1 = - (256/N) \pi^3
[\zeta(3/2)]^{-4/3} \kappa\;\;\;,
\label{c1}
\end{equation}
where $g \kappa= \Delta
\langle \phi^2 \rangle \equiv \langle \phi^2 \rangle_g - \langle \phi^2
\rangle_0$.
The subscripts $g$ and $0$ mean that the field fluctuations
are to be evaluated in the presence and in the absence of interactions
respectively. 

The implementation of LDE within this model is reviewed in  previous
applications \cite{prb,pra,new,braaten}. 
Let us first consider the LDE in the large-$N$
limit extending the relevant $O(2)$ model
to a  $O(N)$-symmetric one. The interest of such
an approximation is that the exact $c_1$ result is known by direct
evaluation \cite{baymN}.
The original perturbation series in this case can be written as
\begin{equation}
\langle \phi^2 \rangle_g^{(k)}= 
- \frac {N \omega}{4\pi} + \frac {N g}{3}
\sum_{i=1}^{k-1} C_i \left ( - \frac { g N}{6 \omega} \right )^i 
\;, \label{exp20}
\end{equation}
where the perturbative coefficients are given by 
\begin{equation}
C_i=  \frac {3}{16 \pi^3} \left ( \frac {1}{8\pi} \right )^i
\int_0^{\infty} dz \frac {z^2}{(z^2+1)(z^2+4)}[A(z)]^i  \;,
\label{Js}
\end{equation}
with

\begin{equation}
A(z) = \frac {2}{z} \arctan \frac {z}{2} \;,
\end{equation}
and $z=k/\omega$, and can be calculated analytically to arbitrary
precision using, e.g., {\it Mathematica} \cite{mathematica}
 numerical integration.
By applying the generalized LDE procedure as described above
in Eqs. (\ref{subst2}), (\ref{subst3}), etc\ldots up to
order $\delta^{9}$, one obtains the  results shown in 
Table III  ($c_{1,IPMS}$), together with
those obtained by the standard PMS optimization procedure, the latter for 
the best converging family of solutions $c_{1,bestPMS}$ (see Ref. \cite{new}
for details on the latter).
\begin{table}[htb]
\begin{center}
\caption{Improved-PMS (IPMS)
versus standard results for the large-$N$ BEC $\Delta T_c$
at different orders $k$. $c_{1,exact}=
2.32847\ldots$} 
\begin{tabular}{|c||c|c||c|}
\hline
$ k $ &  $c_{1,IPMS}$ & $ a $& $c_{1,best PMS}$  \\ \hline\hline
2   &  2.163 &      & 2.163  \\
\hline
3   & 1.85 & 1.44     & $1.88 \pm 0.17 I$  \\
\hline
4  &  1.93 &  1.69      & $1.96$                  \\
\hline
5 &  2.00 &  1.78  &   $1.91 \pm 0.03 I$     \\
\hline
6 &  2.04  & 1.83 &  $1.94$                       \\
\hline
7 &  2.08  & 1.89 &  $1.93 \pm 0.0015I$    \\
\hline
8 &  2.10  & 1.88 &  $1.935$     \\
\hline
9 &  2.12 & 1.89 &   $1.95$     \\
\hline
10 & 2.14 & 1.91   &   $1.95 \pm 0.03I$   \\
\hline
\end{tabular}
\end{center}
\end{table}

As illustrated and discussed in details in \cite{new}, the 
convergence of the standard LDE-PMS method when 
applied to the exact large-$N$ series  Eq. (\ref{exp20}) is actually
very slow (leaving apart the accidentally good result
at first order), so that only at very large 
perturbative orders $\sim 50$ does the optimized
values approach reasonably close to the exact large-$N$ value, 
$c_1 \simeq 2.32847..$. Thus the advantage of the improved PMS
approach is not obvious in this case, since it is evidently less
algebraically tractable than the standard method 
at very large perturbative orders. Nevertheless,
one can see from the improved-PMS results in Table III that 
the new method performs
much better, giving always a unique real solution and 
seemingly converging much
faster towards the expected result.       
In fact, as motivated in \cite{new}, in order to study the
eventual LDE convergence properties of the large-$N$ 
expression of $c_1$,   
it is more instructive to consider an approximated form of the
corresponding large-$N$ perturbative series which takes into account
only the relevant infrared limit of the auxiliary field propagator
(see Ref.  \cite{new} for more details). This
has the advantage of
giving a simple geometric series with exact coefficients: 
\begin{equation}
\langle  \phi^2 \rangle_{IR}^{(k)} = -\frac {N\omega}{4\pi}
+ \frac{ g \; N }{3} \sum_{i=1}^{k-1} G_i
\left (- \frac { g \, N}{6\omega} \right )^i \;\;,
\label{serIR}
\end{equation}
where $G_i \equiv [(64\pi^2)(8\pi)^i]^{-1}$,
such that the straightforward 
resummation of (\ref{serIR}) is well defined
for the relevant limit $\omega \to 0$, which
accordingly can be reached smoothly in contrast
with the genuine large-$N$ series Eq. (\ref{exp20}). The   
alternative method results applied 
on the original series (\ref{serIR}) 
are shown in Table IV, together with
those obtained by the standard PMS optimization procedure, the latter for 
the best converging family of solutions.

\begin{table}[htb]
\begin{center}
\caption{Improved-PMS (IPMS) 
versus standard results for the large-$N$ BEC $\Delta T_c$
at different orders $k$. $c_{1,exact}=
2.32847\ldots$} 
\begin{tabular}{|c||c|c||c|}
\hline
$ k $ &  $c_{1,IPMS}$ & $ a $& $c_{1,best PMS}$  \\ \hline\hline
2   &  2.852 &       & 2.852  \\
\hline
3  &  2.36771 & 1.49     & $2.444 \pm 0.276I$                     \\
\hline
4 &  2.34451 &  1.53 &   $2.244 \pm 0.20 I$     \\
\hline
5 &  2.33686  & 1.56 &  $2.397 \pm 0.079I$                       \\
\hline
6 &  2.32847  & 2.0 &  $2.333 \pm 0.08I$    \\
\hline
7 &  2.32847  & 2.0 &  $2.298\pm 0.06I$     \\
\hline
8 &  2.32847  & 2.0 &   $2.342\pm 0.04I$     \\
\hline
9 & 2.32847  & 2.0   &   $2.324\pm 0.036I$   \\
\hline
\end{tabular}
\end{center}
\end{table}


In that case, the convergence of the standard PMS method
is faster in comparison of Table III, but the improved PMS
method performs even better:
starting from order six and beyond, the exact result is
always obtained as the unique physically acceptable solution.
It is interesting to trace what
happens in more details. In fact, the solutions found at 
LDE orders $k \ge 6$ by  applying
the procedure is $a=2$, $b=c=d=\ldots=0$, which is easily seen by comparing
with Eqs. (\ref{subst2}), (\ref{subst3}) to correspond to a basic interpolation
of the form  
\begin{equation}
\omega \to \omega (1-2\,\delta +\delta^2 +0\:)^{1/2} = 
\omega (1-\delta) \;\;,
\label{subst4}
\end{equation}
Then, applying
the substitution (\ref{subst4}) on any geometric 
series instead of the standard
LDE   substitution (\ref{subst}) canonical for a scalar theory,
one can easily see from simple algebraic properties that the 
improved PMS 
solution always reproduces the exact solution,
and this at any arbitrary LDE order $k \ge 2$. 
This case exhibits a spectacular improvement over the standard
LDE results, which converges only rather slowly and with non-zero
(albeit small) imaginary parts, a source of much frustration.
{\em A posteriori}, there is nothing particularly 
remarkable in this
result which is essentially an algebraic accident of the 
non-canonical substitution Eq. (\ref{subst4}) followed
by the LDE when performed on a simple geometric
series. However, what is perhaps more interesting is that
our improved-PMS procedure is in that way guessing 
a more appropriate value of the rescaling power 
within the LDE substitution Ansatz Eq. (\ref{subst1}):
Alternatively we could have parametrized the 
basic LDE substitution
according to    
\begin{equation}
\omega \to \omega (1-\delta)^\gamma \;\;,
\label{expcrit}
\end{equation}
with an arbitrary power $\gamma$ to begin with, and then look for 
the best value of $\gamma$ such that the LDE series converges
faster
towards the exact result, which is clearly the case for $\gamma=1$
in this large-$N$ case. 
Now, it is worth noting that  considering an 
arbitrary power coefficient $\gamma$ according to Eq. (\ref{expcrit})
when applied to the BEC series turns out in practice to be essentially 
equivalent to modifying the simplest LDE substitution formula Eq. 
(\ref{subst}) by introducing
the relevant critical exponent:
\begin{equation}
\omega \to \omega (1-\delta)^{1/\omega^\prime} \;\;,
\label{expcrit2}
\end{equation}
where
$\omega^\prime =
2\,\Omega/(2-\eta)$, $\eta$ is the anomalous dimension
of the critical propagator $\sim 1/p^{2-\eta}$
and $\Omega \equiv \beta^\prime(g_c)$,
$g_c$ being the critical coupling (see e.g. Ref. \cite{zustin}).  
This renormalization group inspired  modification 
Eq. (\ref{expcrit2}) of 
the standard LDE Eq. (\ref{subst1}) is indeed the approach 
followed e.g. in Refs. \cite{kleinert,
kastening2}, where numerical values of $\omega^\prime$ as obtained
by different methods (including the variational perturbation theory
\cite{kleinert2}) are used in Ansatz (\ref{expcrit2})
prior to an (otherwise standard)
PMS optimization.   
Now in our case, a major difference is that the relevant
exponent $\omega^\prime \simeq 2/a$ 
is simply guessed by the generalized optimization procedure, 
at the same time as obtaining as optimized solution
 the relevant physical quantity $c_1$. More precisely,
 the correct exact large-$N$ value
$\omega^\prime =1$,
$\Omega =1$ (see e.g. Ref. \cite{zustin}) is clearly guessed by
our improved PMS procedure, with 
$\omega^\prime =2/a$, at least for the 
infrared approximated large-$N$ case 
as illustrated in Table IV. Note also that the values of the parameter
$a$ at successive perturbative orders
are also clearly approaching quite closely, though more slowly,  
the correct critical value 
when considering the genuine large-$N$ series, as illustrated
 in Table III (noting however that
in this case the other parameters $b$, etc\ldots are small with 
respect to $a$ but not strictly zero). \\

We will consider now mainly the physically relevant $N=2$ case, 
as well as the case $N=1$ and $N=4$ for comparison with other
available lattice simulation and analytical results. 
We will see that many of the previous results for the large-$N$ series
generalize, at least qualitatively, to this less trivial case.
For $N=2$, 
the quantity $\langle \phi^2 \rangle_g^{(k)}$  has been first
evaluated, up to order $\delta^4$, in Ref. \cite {pra}. Recently, higher order
terms have been evaluated by Kastening \cite
{kastening2} so that, to order $\delta^6$, the perturbative series can be
written as
\begin{equation}
\langle \phi^2 \rangle_g^{(6)}= - \frac {N \omega}{4\pi} + g
\sum_{i=1}^{5} K_i \left ( - \frac {g}{\omega} \right )^i   \;,
\label{expNF}
\end{equation}
where the coefficients are given by\footnote{Throughout this paper we use 
the perturbative loop coefficients results of  Ref. \cite{kastening2} to
evaluate all these coefficients, since in
particular the latter are obtained as much as possible from exact
integrals. There are therefore some very small differences in the
lowest orders numerical results with respect 
to some previous analysis \cite{pra,new}, since the latter used numerical
integration,
not always very precise.  
Note also a trivial difference of normalization
with respect to  Ref. \cite{kastening2}
in our defining series, Eq. (\ref{expNF}).}
 $K_1=3.22174\times10^{-5}$,
$K_2=1.51792 \times 10^{-6}$, $K_3=9.66512 \times 10^{-8}$,
$K_4\simeq 7.51366 \times 10^{-9}$,
and $K_5 \simeq 6.7493 \times 10^{-10}$. The results
of our alternative procedure are shown in Table V, together
with the results from applying simple Pad\'e approximants 
similar to
the ones discussed above for the oscillator, 
i.e. defining a new perturbative series similar
to Eq. (\ref{padseries}) but where the $E_i$ are now replaced
by the values of $c_1$ in Table V at successive orders $k$.
One can notice that at orders
$k=2,3$ and $4$ the IPMS results quickly oscillate around the MC results.
 From order $\delta^4$ onwards the oscillation is reduced drastically and the
IPMS procedure generates stable results which agree remarkably well with the
lattice results. Furthermore, we also indicated in Table
V the corresponding values obtained  at each order for
the parameter $a$, whereas as already indicated,
in the generalized LDE substitutions Eq. (\ref{subst3}), etc\ldots
at higher orders the remaining parameters $b$, etc are numerically
smaller. 
Accordingly, to first order approximation,
\begin{eqnarray}
\omega & \to  & \omega \:
[1-a\,\delta -(1-a)\,\delta^2 +b\,\delta^2(1- \delta)+
\cdots]^{1/2}\; \nonumber \\
 & \simeq  & \omega \:(1-\delta)^{a/2}\;,
\end{eqnarray}
so that in a rough approximation one has 
$\omega^\prime \simeq 2/a$. Taking the latter approximate
relation at face value would give e.g. for the successive
orders considered in Table V: $\omega^\prime \sim $
 1.03,  0.71, 0.73, 0.71, which compare not badly with
the reported values of $\omega^\prime = 0.8 \pm 0.04$
\cite{zustin,kleinert2} obtained numerically by other
 methods. Note, however, that
the present method does not pretend at this stage
to accurately predict in that way the relevant critical exponent
$\omega^\prime$, $\Omega$ etc. Indeed, in this $N=2$ 
 case the other parameters $b$, etc\ldots are small with 
respect to $a$ but not strictly negligible. Nevertheless,
though the essential motivation of our approach introducing
more interpolation parameters in Eq. (\ref{subst3})
is  to get rid of the unwanted complex
optimization solutions, it is interestingly connected with the 
parametrization of corrections to scaling, the leading correction
being correctly described according to Eq. (\ref{expcrit2})
(see e.g. Ref. \cite{kleinert,kastening2} for further motivation
of Eq. (\ref{expcrit2})).  
However, to be complete one should note that, 
in contrast with the previous
oscillator and large-$N$ BEC series, the real IPMS solutions in Table V
are not always unique: actually, at orders $k =4$ and $k=6$
there appear a second real (but negative) 
solution. But these extra solutions can be immediately eliminated as 
they correspond to largely unreasonable values of the
interpolation parameters: for instance for $k=4$ 
the extra real solution is: $a \simeq -13.9$, $b \simeq 55.5$, and 
$\tilde \omega \simeq -0.02$ which gives $c_1 \simeq -55.4$,
while we can 
expect consistent values of these parameters to be reasonably close
to their corresponding large-$N$ values, i.e. typically $a$ should
be not too far from its large-$N$ value $a=2$, as also
supported from the results of 
the above mentioned leading corrections to scaling analysis. 

At the same time, the results
from the standard LDE-PMS optimization method  shown in Table V, 
also  oscillate somehow
around the MC result, as anticipated
in Ref. \cite {braaten}, but the convergence is not at all obvious:
after approaching rather closely the lattice
result at  order $\delta^4$, the results of the next two
higher orders depart sensibly from the lattice one. 
In the absence
of any indication on the higher orders, we can hardly speculate
but, by comparison with
the results of Ref. \cite{new} for the large-$N$
case, as summarized in Tables III and IV, it appears  
possible that the standard LDE ultimately converges towards the same
result, but in a much slower way than our improved method given
in the left column of Table V.  
Moreover, the latter standard PMS results will look much 
less attractive if
one recalls that they have been selected among 
several possible complex results with large imaginary
parts \cite{new}, requiring an 
extra selection criteria\footnote{More precisely, 
we selected\cite{new} among the 
different LDE-PMS real or complex solutions the ones having  
the smallest $Re[\tilde \omega] > 0$ value of the optimized mass,
motivated by the fact that the exact
solution would correspond to $ \omega \to 0$.}, in contrast to the 
improved version.

One can also note the remarkable results of the Pad\'e approximants
based on the improved PMS results
at different orders, even for the lowest order one $P_{[1,1]}$, though
it only uses the second and third order IPMS
results. This is certainly not coincidental and should be mainly 
attributed to the oscillatory property of the IPMS results. Indeed,
the very same Pad\'e approximants  using instead the (real parts of)
the standard PMS results of the rightmost column of Table V are
far away from any reasonable result: this would give e.g. 
$P_{[1,1]} =4.26$ and $P_{[2,2]}=3.66$.\\
\begin{table}[htb]
\begin{center}
\caption{Improved-PMS ($IPMS$)
versus standard ($PMS$) results for $c_1$ at $N=2 $ in the  BEC case  at
different orders $k$. Also shown are the
Pad\'e approximants (PA) resummation results applied to  the
improved-PMS cases. The lattice results are $c_1 = 1.29 \pm 0.05$,
$1.32\pm 0.02$.} 
\begin{tabular}{|c||c|c|c||c|}
\hline
$ k $ &  $c_{1,IPMS}$ &$a$ & $ PA(c_{1,IPMS})$ & $c_{1,PMS}$  \\ \hline\hline
2   &  3.06 &  &   & 3.06     \\
\hline
3   &  0.98 & 1.95 &     & $2.45\pm 1.66I$   \\
\hline
4  &  1.426 & 2.81 & $P_{[1,1]}=1.347$  & $1.53\pm 2.32I$     \\
\hline
5 &  1.247 & 2.75  & $P_{[1,2]}=1.283$ ;
$P_{[2,1]}=1.298$  & $0.76  \pm 2.53I$   \\
\hline
6 &  1.300 & 2.83 & $P_{[2,2]}=1.286$  & $2.40  \pm 1.69I$    \\
\hline
\end{tabular}
\end{center}
\end{table}

Next, we consider the completely similar calculation 
of the coefficient $c_1$ as defined in Eq. (\ref {c1}) but for
the cases $N=1$ and $N=4$, respectively, for which
lattice calculations of $c_1$ have been recently performed
in Ref. \cite{lattn14}. 
The original perturbative series reads like Eq. (\ref{expNF})
with now the relevant coefficients $K_i$ given 
 by
$K_1=1.208097 \times10^{-5}$,
$K_2=5.1229853 \times 10^{-7}$, $K_3=2.9645225 \times 10^{-8}$,
$K_4\simeq 2.1084956 \times 10^{-9}$,
and $K_5 \simeq 1.7420267 \times 10^{-10}$, and
$K_1=9.6647794 \times10^{-5}$,
$K_2=5.4645177 \times 10^{-6}$, $K_3=4.09584576 \times 10^{-7}$,
$K_4\simeq 3.70261198 \times 10^{-8}$,
and $K_5 \simeq 3.8333358 \times 10^{-9}$, respectively
for $N=1$ and $N=4$.

Our results are given in Tables 
VI and VII respectively for $N=1$ and $N=4$. As one can see, the
results from our alternative method are again showing very
good convergence properties and approaching very closely the
lattice results. The Pad\'e approximants
are also in excellent agreement with the latter. 
In addition,
the corresponding values obtained for the parameter
$a$, as related in first approximation to the critical exponent 
$\omega^\prime$ in Eq.~(\ref{expcrit}), appear 
qualitatively very consistent to
known results 
for $\omega^\prime$ for $N=1$ and $N=4$\cite{zustin,kleinert2}. 
On the other hand, the results
from the standard LDE-PMS method show a behavior very similar
to the case $N=2$: after approaching not too far from the lattice
result at fourth order (but with very large imaginary parts), 
the results from higher orders
are not satisfactory, with trends very similar to the case $N=2$.


\begin{table}[htb]
\begin{center}
\caption{Same as Table V (improved versus standard) PMS 
results but for $N=1 $   at
different orders $k$. The lattice result is $  1.09 \pm 0.09$.}
\begin{tabular}{|c||c|c|c||c|}
\hline
$ k $ &  $c_{1,IPMS}$ & $a$& $ PA(c_{1,IPMS})$ & $c_{1,PMS}$   \\ \hline\hline
2   &  2.65 &   &  & 2.65     \\
\hline
3   &  0.817 & 1.95  &   & $2.12\pm 1.47I$   \\
\hline
4  &  1.237 & 2.88 &  $P_{[1,1]}=1.159$  & $1.31\pm 2.06I$     \\
\hline
5 &  1.047 & 2.77& $P_{[1,2]}=1.086$;  $P_{[2,1]}=1.106$ 
& $0.62  \pm 2.29I$   \\
\hline
6 &  1.114 &2.90 & $P_{[2,2]}=1.095$  & $2.11  \pm 1.54I$    \\
\hline
\end{tabular}
\end{center}
\end{table}
\begin{table}[htb]
\begin{center}
\caption{Same as Table V (improved versus standard) PMS results 
but for  $N=4 $  at
different orders $k$. The lattice result is $ 1.59 \pm 0.10$.}

\begin{tabular}{|c||c|c|c||c|}
\hline
$ k $ &  $c_{1,IPMS}$& $a$ & $ PA(c_{1,IPMS})$ & $c_{1,PMS}$  \\ \hline\hline
2   &  3.75 &  &  & 3.75      \\
\hline
3   &  1.222 & 1.94 &  & $2.99\pm 1.99I$    \\
\hline
4  &  1.665 &2.68 &  $P_{[1,1]}=1.589$   & $1.90\pm 2.74I$     \\
\hline
5 &  1.524 & 2.66 & $P_{[1,2]}=1.550$;  $P_{[2,1]}=1.558$
& $0.99  \pm 3.01I$   \\
\hline
6 &  1.556 & 2.74  & $P_{[2,2]}=1.549$  & $2.90  \pm 1.98I$    \\
\hline
\end{tabular}
\end{center}
\end{table}

We finally apply our alternative method to the other non-perturbative
relevant coefficient $c_2^{\prime \prime}$ in the defining Eq. (\ref{Tcexp}).
This coefficient, which is not generally considered by most
authors working on the BEC $\Delta T_c$ problem,  was first evaluated with
lattice simulations. To our knowledge,
its sole analytical  evaluation made use of the standard 
LDE-PMS \cite{pra,new} up to
order $\delta^4$ (five loops). This $O(a^2 n^{2/3})$ 
coefficient appearing in the $\Delta T_c$ expansion,
Eq. (\ref {Tcexp}), can be written as \cite {second}
\begin{eqnarray}
c_2^{\prime\prime} &=& - \frac{2}{3} [\zeta (3/2)]^{-5/3} 
b_2^{\prime \prime} +
\frac {7}{9}
[\zeta (3/2)]^{-8/3} (192 \pi^3 \kappa)^2 \nonumber \\  &+ &\frac{64
\pi}{9} \zeta (1/2) [\zeta(3/2)]^{-5/3}
\ln \zeta (3/2)   \;,
\label{c2}
\end{eqnarray}
where $b_2^{\prime \prime}$ is
\begin{eqnarray}
b_2^{\prime \prime} &= 32 \pi \bigl\{ \left[ \frac{1}{2} \ln (128 \pi^3) +
\frac{1}{2} - 72 \pi^2 {\cal R} - 96 \pi^2 \kappa \right]  \nonumber \\
&  \times \zeta(1/2)
 + \frac {\sqrt {\pi}}{2} - K_2 - \frac {\ln 2}{2 \sqrt {\pi}}\left [
\zeta(1/2) \right ]^2  \bigr\}  \; ,
\label{b2primeprime}
\end{eqnarray}
with $K_2= -0.13508335373$ and $g^2
{\cal R}= r_c$ where $r_c=-\Sigma(0)$ (Hugenholtz-Pines theorem).
As before, the quantity $\kappa$ is easily obtained from Eq.
(\ref {expNF}) whereas  ${\cal R}$ can be obtained directly from the
perturbative evaluation of  $\Sigma(0)$. We refer the interested reader to
Ref. \cite {pra} for the details and subtleties 
associated with this type of
evaluation. Thanks to the recent availability of improved six-loop results
\cite{kastening1} one obtains , at $N=2$,

\begin{eqnarray}
\delta r_c^{(5)}&=& - \Sigma_{\rm ren}^{(5)}(0)=  \frac {g \omega}{6\pi}
+ g^2 A_2 \left [ \ln
\left ( \frac{M}{\omega} \right ) - 0.59775 \right ] \nonumber \\
&+& \omega^2 \sum_{i=3}^5 \left ( \frac {- g }{\omega} \right )^i A_i \;\;,
\label{rcNF}
\end{eqnarray}
where $M$ is an arbitrary mass scale introduced by dimensional
regularization and $A_2=1.40724 \times 10^{-3}$, $A_3=8.50888 \times
10^{-5}$, $A_4=3.57259 \times 10^{-6}$ and $A_5= 2.25332 \times 10^{-7}$.
As discussed in Ref. \cite {pra} it is interesting to note that the 
optimized $\tilde \omega$ defined by the standard PMS, or as well
by the IPMS introduced in the present work, are scale independent to any
order in $\delta$ so that $\tilde \omega$ is only $g$ dependent. Note
also that, contrary to $\Delta \langle \phi^2 \rangle$, $r_c$ is a divergent
quantity. In this case, the optimization procedure must be implemented after
renormalization as advocated in Ref. \cite {marcus1}.
Finally, let us point out that the potential technical difficulty associated
with the evaluation of $r_c$ stems from the fact that, while
$\Delta \langle \phi^2 \rangle$ depends on the (finite)  difference
$\Sigma(p)-\Sigma(0)$, the former depends on the (divergent) $\Sigma(0)$ only.
Here, we had access \cite {private} to values of $\Sigma(0)$ up to
order $\delta^5$ (six loops) only. Unfortunately, at order $\delta^6$ only the
results for the joint $\Sigma(p)-\Sigma(0)$ contribution is known \cite
{kastening1}. \\
The results of the alternative method are shown in Table VIII below for $M=g/3$
which was the value chosen in the lattice evaluations \cite {second}. As
one can see, the results of our alternative method show again a remarkable
agreement with the numerical lattice result
up to the fifth order (six loop). Concerning the results
of the standard PMS method, as shown in the right column of Table VIII,
their behavior is again very similar to the ones for the $c_1$ coefficient
above in Tables V--VII: after approaching very closely the lattice
result at fourth (five-loop) order, the six-loop result starts to decrease
a bit further. \\
\begin{table}[htb]
\begin{center}
\caption{Improved-PMS results for $c_2^{\prime \prime}$ in the $N=2$ BEC case
at different orders $k$. The lattice result is $ c_2^{\prime \prime}=
75.7 \pm 0.4$.}
\begin{tabular}{|c||c|c||c|}
\hline
$ k $ &  $c_{2,IPMS}^{\prime \prime}$ & Pad\'e($c_{2,IPMS}^{\prime \prime}$)
& $c_{2,best PMS}^{\prime \prime}$ 
  \\ \hline\hline
2   & 101.2 &       & 101.2      \\
\hline
3   &  69.83 &    & $94.2 \pm 31.2 I$    \\
\hline
4  &  77.90 &    $P_{[1,1]}=  76.25$     & $75.0 \pm 41.1 I$     \\
\hline
5 &  73.46 & $P_{[1,2]}=79.31$; $P_{[2,1]}=75.04$  &  $60.4 \pm 41.2 I$     \\
\hline
\end{tabular}
\end{center}
\end{table}

\section{conclusion} 

We have introduced a conceptually rather
simple  generalization of the
optimized (or variationally improved) perturbation method,
which considerably reduces the number of irrelevant optimization
solutions at each  perturbative order. We have applied
this variant of the PMS method to the calculation of certain
non-perturbative quantities in the simple quantum mechanical
oscillator as well as the more challenging BEC critical 
temperature determination. 

Just like the usual LDE/PMS procedures, it is a recipe without a formal
general justification.  For quantities described by a
perturbative series which is originally factorially divergent, such
as the series relevant
for the oscillator, we find that this new procedure, though it helps
in selecting only a few (if not always uniquely) among the
many possible complex solutions, does not exhibit
obvious convergence improvement behavior, in contrast
with the standard
method whose rigorous convergence is established in simple cases. 
Our experience from several 
toy series as well as the physical
systems considered in this paper seems to point to the following behaviors
which remain somewhat mysterious to us: Compared to 
the usual LDE/PMS procedures, it seems much
superior conceptually 
when applied 
to  series with finite radius of convergence or at least
less divergent series,
like the ones relevant to the BEC critical
temperature.
 When one deals
with an asymptotic expansion (with zero radius of convergence
but alternating signs) typical of those appearing in perturbative 
calculations for physical systems, it seems
that the recipe breaks down at high order, but that at low order it
gives excellent numerical results, provided that the asymptotic expansion
is not too different in its first few orders from
a series with finite radius of convergence
(Examples are provided by finite-$N$ versus large-$N$ theories).\\
Moreover, quite interestingly the method appears
in non-trivial cases to estimate numerically (by optimization)  
essentially correct values of the leading corrections to scaling behavior,
since the introduction of more interpolation parameters in the
LDE Anstaz is, to a first approximation, equivalent to introducing
the relevant critical exponent. 
Finally, our numerical results
obtained for the BEC critical temperature when using the latest available
perturbative
information are in excellent agreement with the numerical
lattice simulations results for the all available finite $N$ cases 
(and also for the large-$N$ case as 
compared to the exact analytic result).

\acknowledgments

A.N. is partially supported by EU under contract EUCLID
HPRN-CT-2002-00325. 
M.B.P. is partially supported by CNPq-Brazil.
We thank Boris Kastening for useful communication relevant to
the six loop evaluation of $r_c$.


\begin{thebibliography}{99}

\bibitem{delta} 
V. Yukalov, Teor. Mat. Fiz. {\bf 28}, 92 (1976);
W.E. Caswell, Ann. Phys. (N.Y) {\bf 123}, 153 (1979);
I.G.  Halliday and P. Suranyi, Phys. Lett. 
 {\bf B85}, 421 (1979);
J. Killinbeck, J. Phys.  {\bf A14}, 1005 (1981);
R.P. Feynman and H. Kleinert, Phys. Rev. {\bf A34}, 5080 (1986);
A. Okopinska, Phys. Rev.  {\bf D35}, 1835 (1987);
A. Duncan and M. Moshe, Phys. Lett. {\bf B215}, 352 (1988);
H.F. Jones and M. Moshe, Phys. Lett. {\bf B234}, 492 (1990);
A. Neveu, Nucl. Phys. (Proc. Suppl.) {\bf B18}, 242 (1990);
V. Yukalov, J. Math. Phys {\bf 32}, 1235 (1991);
S. Gandhi, H.F. Jones and M. Pinto, Nucl. Phys. {\bf B359}, 429 (1991);
C.~M. Bender et al., Phys. Rev. {\bf D45}, 1248 (1992);
S. Gandhi and M. Pinto, Phys. Rev. {\bf D46}, 2570 (1992);
H. Yamada, Z. Phys. {\bf C59}, 67 (1993);
K.G. Klimenko, Z. Phys. {\bf C60}, 677 (1993);
A.N. Sissakian, I.L. Solovtsov and O.P. Solovtsova, 
Phys. Lett. {\bf B321}, 381 (1994);
H. Kleinert, Phys. Rev. {\bf D57}, 2264 (1998); Phys. Lett. {\bf B434}, 74
(1998); see also {\em Critical Properties of $\phi^4$-Theories},
H. Kleinert and V. Schulte-Frohlinde, World Scientific (2001),
chap.~19 for a recent review.

\bibitem{pms} P. M. Stevenson, Phys. Rev. {\bf D23}, 2916 (1981);
Nucl. Phys. {\bf B203}, 472 (1982).

\bibitem{odm} R. Seznec and J. Zinn-Justin, J. Math. 
 Phys. {\bf 20}, 1398 (1979); J.C. Le Guillou and J. Zinn-Justin,
Ann. Phys. {\bf 147}, 57 (1983).

\bibitem{ao} C.M. Bender and T.T. Wu, Phys. Rev. {\bf 184}, 1231 (1969);
Phys. Rev. {\bf D7}, 1620 (1973).

\bibitem{deltaconv} R. Guida, K. Konishi and H. Suzuki, Ann. Phys. 241
(1995) 152;  Annals Phys. {\bf 249}, 109 (1996).
\bibitem{deltac} A. Duncan and H.F. Jones, Phys. Rev. {\bf D47}, 2560 (1993);
 C.M. Bender, A. Duncan and H.F. Jones, Phys. Rev. {\bf D49}, 4219 (1994);
B.~Bellet, P.~Garcia and A.~Neveu, Int.J.Mod.Phys. {\bf A11}, 5587 (1996);
{\it ibid.} {\bf A11}, 5607 (1996);   C. Arvanitis, H.F. Jones and C. Parker,
Phys. Rev. {\bf D52}, 3704 (1995) 3704;  H. Kleinert, Phys. Rev. Lett. {\bf 75},
2787 (1995); W. Janke and H. Kleinert, Phys. Lett. {\bf A206}, 283 (1995).

\bibitem{gn2} C.~Arvanitis, F.~Geniet, M. Iacomi, J.-L.~Kneur and A.~Neveu,
Int.J.Mod.Phys. {\bf A12}, 3307 (1997).

\bibitem{marcus1}M. B. Pinto and R. O. Ramos, Phys. Rev. {\bf D60 },
105005 (1999); {\it ibid.} {\bf D61}, 125016 (2000).

\bibitem{pra}F. F. S. Cruz, M. B. Pinto, R. O. Ramos and P. Sena,
Phys. Rev. {\bf A65}, 053613 (2002).

\bibitem{new}J.-L. Kneur, M. B. Pinto and R. O. Ramos,
Phys. Rev. Lett. {\bf 89}, 210403 (2002); Phys. Rev. {\bf A68}, 043615 (2003).


\bibitem{braaten} E. Braaten and E. Radescu,  
Phys. Rev. Lett. {\bf 89}, 271602 (2002);  cond-mat/0206186.\\

\bibitem{zustin}J. Zinn-Justin, {\it Quantum Field Theory and Critical 
Phenomena} (Oxford University Press, 1996).

\bibitem{second} P. Arnold, G. Moore and B. Tom\'{a}sik, 
Phys. Rev. {\bf A65}, 013606 (2002).

\bibitem{baymprl} G. Baym, {\it et. al}, Phys. Rev. Lett. {\bf 83}, 1703 (1999).

\bibitem{baymN} G. Baym, J.-P. Blaizot and J. Zinn-Justin, Europhys.
Lett. {\bf 49}, 150 (2000).

\bibitem{arnold1} P. Arnold and B. Tom\'{a}sik, Phys. Rev. {\bf A62},
063604 (2000).

\bibitem{mathematica} Mathematica version 3.0, S. Wolfram Company.

\bibitem{prb}F. F. S. Cruz, M. B. Pinto and R. O. Ramos,
Phys. Rev. {\bf B64}, 014515 (2001); Laser Phys. {\bf 12}, 203 (2002).

\bibitem{russos} V.A.  Kashurnikov, N.V. Prokof'ev and B.V. Svistunov,
Phys. Rev. Lett. {\bf 87}, 120402 (2001).

\bibitem{arnold2} P. Arnold and G. Moore, Phys. Rev. Lett. {\bf 87},
120401 (2001); Phys. Rev. {\bf E64}, 066113 (2001).

\bibitem{kastening2} B. Kastening, e-Print Archive: cond-mat/0309060.

\bibitem{kleinert} H. Kleinert, e-Print Archive: cond-mat/0210162.
\bibitem{kleinert2} H. Kleinert, Phys. Rev. {\bf D60}, 085001 (1999).

\bibitem{kastening1} B. Kastening, Phys. Rev. {\bf A68}, 061601 (2003).

\bibitem{private} B. Kastening, private
communication.

\bibitem{lattn14} X.-P. Sun, Phys. Rev. {\bf E67}, 066702 (2003).

\end{thebibliography}
\end{document}